\documentclass[a4paper,11pt]{article}
\usepackage{pos}
\usepackage{xspace}

\title{Intrinsic $k_T$ and soft gluons in Monte Carlo event generators}
\ShortTitle{Intrinsic-$k_T$ and soft gluons in Monte Carlo event generators}

\author*[a]{L. Moureaux}
\author[b]{A. Lelek}
\author[c]{F. Hautmann}
\author[d]{L. Favart}

\affiliation[a]{
Universit{\" a}t Hamburg,
    Institut f{\" u}r Experimentalphysik,
    Luruper Chaussee 149, Hamburg, Germany}

\affiliation[b]{Elementaire Deeltjes Fysica, 
Universiteit Antwerpen, B 2020 Antwerpen, Belgium}

\affiliation[c]{University of Oxford,  
Oxford OX1 3PU, United Kingdom}

\affiliation[d]{IIHE,    Université libre de Bruxelles,
    Boulevard du Triomphe, 2, 1050 Brussels, Belgium}

\emailAdd{louis.moureaux@cern.ch}

\abstract{
Experimental measurements of the 
transverse momentum of Drell-Yan lepton pairs   
are sensitive to non-perturbative physics associated with the
intrinsic parton transverse momentum $k_T$.
We discuss     recent determinations of intrinsic $k_T$  
 in the context of   transverse momentum dependent (TMD) 
 parton branching  calculations and  collinear parton-shower Monte Carlo 
 generators.  We illustrate 
 the influence of the soft-gluon resolution scale and   the           
 non-perturbative Sudakov region on the intrinsic $k_T$ 
 extraction.  We emphasize the relevance of the correct treatment of  
 correlated uncertainties between different transverse momentum bins
 in TMD fits and present an 
 application to the determination of the intrinsic $k_T$ in the
 forward rapidity region.  
   }

\FullConference{The European Physical Society Conference on High Energy Physics EPS-HEP2025\\
 7-11 July 2025\\
Marseille, France\\
}

\begin{document}
\maketitle

\paragraph{Introduction}

 The intrinsic 
transverse momentum $k_T$ of partons in  hadronic states at high energies 
is one of the most important contributions to non-perturbative physics in 
production processes at hadron colliders. 
The intrinsic $k_T$ 
measures transverse degrees of freedom 
of the strong interactions  
for low mass 
scales in the QCD evolution~\cite{jccbook}.  
It is taken into account in 
parton-shower Monte Carlo event 
generators~\cite{Buckley:2019kjt} which are used   
 in analyses of experimental data at colliders, 
 as well as  
 in QCD calculations based on 
 factorization  and evolution of 
 transverse momentum dependent (TMD) 
 parton distributions~\cite{Angeles-Martinez:2015sea,Abdulov:2021ivr}. 

\begin{figure}[h]
  \begin{centering}
    \includegraphics[height=0.41\textwidth]{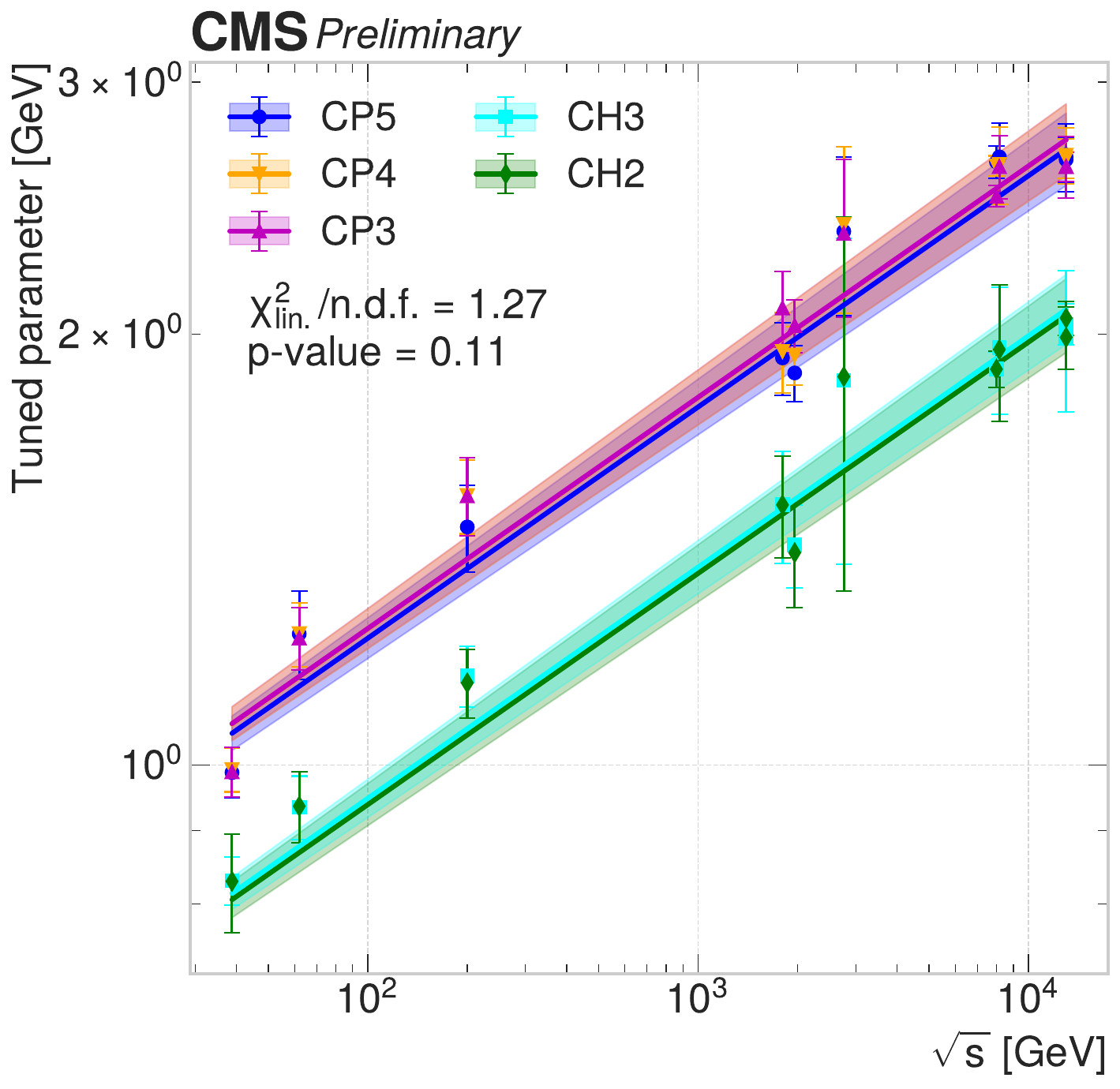}
    \includegraphics[height=0.4\textwidth]{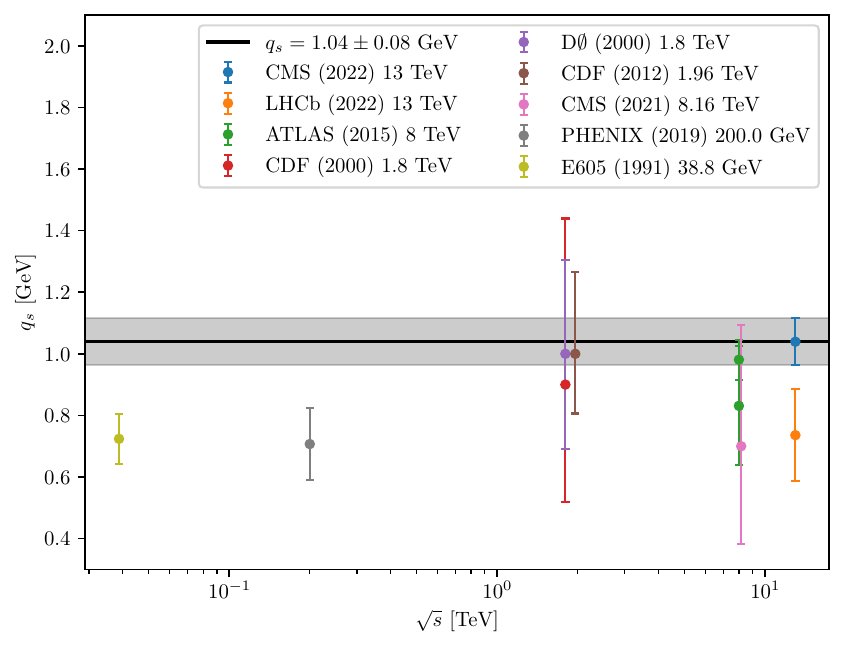}
    \caption{%
        Extraction of intrinsic $k_T$ parameters from measurements of DY transverse
        momentum distributions at varying center-of-mass energies.
        The plot on the left-hand side~\protect\cite{CMS:2024goo} is
        based on the collinear parton-shower Monte Carlo generators
        {\sc Herwig}~\protect\cite{Bellm:2015jjp} and {\sc Pythia}~\protect\cite{Sjostrand:2014zea},
        with underlying-event tunes CH2 and CH3~\protect\cite{CMS:2020dqt} and
        CP3 through CP5~\protect\cite{CMS:2019csb}, respectively.
        The plot on the right-hand side~\protect\cite{Bubanja:2023nrd} is based on the
        PB TMD method, with the horizontal line based on CMS 13\,TeV
        data~\cite{CMS:2022ubq}.
    }
    \label{fig:extract-intrins}
  \end{centering}
\end{figure}

The intrinsic $k_T$ distribution can be extracted from comparisons of theory
predictions with experimental measurements of the transverse momentum $p_T$ of lepton
pairs in the Drell-Yan (DY) process.
This has recently been carried out for the {\sc Herwig}~\cite{Bellm:2015jjp} and {\sc Pythia}~\cite{Sjostrand:2014zea}
collinear parton showers~\cite{CMS:2024goo} and for the parton
branching 
(PB)~\cite{Hautmann:2017fcj,Hautmann:2017xtx,BermudezMartinez:2018fsv,BermudezMartinez:2021lxz}  
formulation of TMD evolution~\cite{Bubanja:2023nrd}.
A good description of the low-$p_T$ experimental data is achieved in both cases.
For parton showers, this necessitates a value of the intrinsic-$k_T$ parameter
that rises steeply with energy (Fig.~\ref{fig:extract-intrins} left).
This is not needed in the PB TMD approach (Fig.~\ref{fig:extract-intrins} right).

In these proceedings, we discuss two aspects of these studies and the
determination of intrinsic $k_T$ parameters.
First, we recall the results of Ref.~\cite{Hautmann:2025fkw},
interpreting the different behaviors seen in Fig.~\ref{fig:extract-intrins}
from the standpoint of including (or not) the evolution of the partonic
transverse momentum in the parton cascade~\cite{Dooling:2012uw}
 through TMD distributions, and  
  in particular the role of QCD emissions 
  beyond the soft-gluon resolution scale~\cite{Hautmann:2019biw}.

The second aspect concerns the value of the intrinsic $k_T$ parameter $q_s$
extracted from the analysis of LHCb measurements~\cite{LHCb:2021huf} shown
in Fig.~\ref{fig:extract-intrins} (right), which appears to be about 30\%
lower than values obtained from the analysis of CMS~\cite{CMS:2022ubq,CMS:2019raw}
and ATLAS~\cite{ATLAS:2019zci} measurements at similar energies.
 We revisit the study of the LHCb dataset~\cite{Favart:2025inprep}
 presenting new results
 which take into account a number of potentially significant 
 factors, including   the contribution of forward 
 rapidities to the transverse momentum distribution, 
 the treatment  of final-state radiation,   the 
 role  of  experimental uncertainties and especially correlations  
 in the extraction of the intrinsic $k_T$ parameters.

All PB TMD results shown in these proceedings use the
\textsc{Cascade}~\cite{CASCADE:2021bxe,CASCADE:2010clj} Monte Carlo generator.
The hard scattering is simulated at next-to-leading order with
\textsc{MadGraph5\_aMC@NLO}~\cite{Alwall:2014hca}, matched to the TMD parton
distributions and initial-state shower as described in
Refs.~\cite{BermudezMartinez:2019anj,BermudezMartinez:2020tys,Abdulhamid:2021xtt,Yang:2022qgk}.
Final-state radiation and hadronization are simulated with
\textsc{Pythia}~6~\cite{Sjostrand:2006za}.

\begin{figure}[h]
    \centering
    \includegraphics[width=0.47\textwidth]{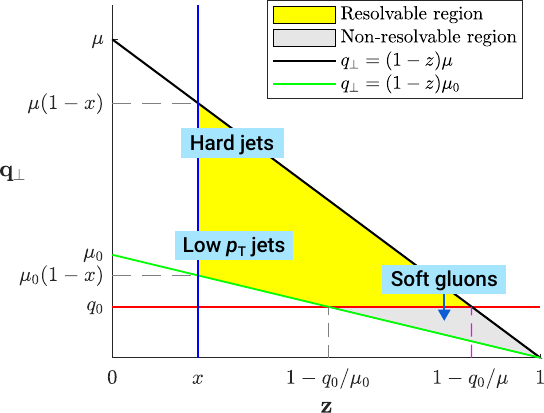}\relax
    \includegraphics[width=0.53\textwidth]{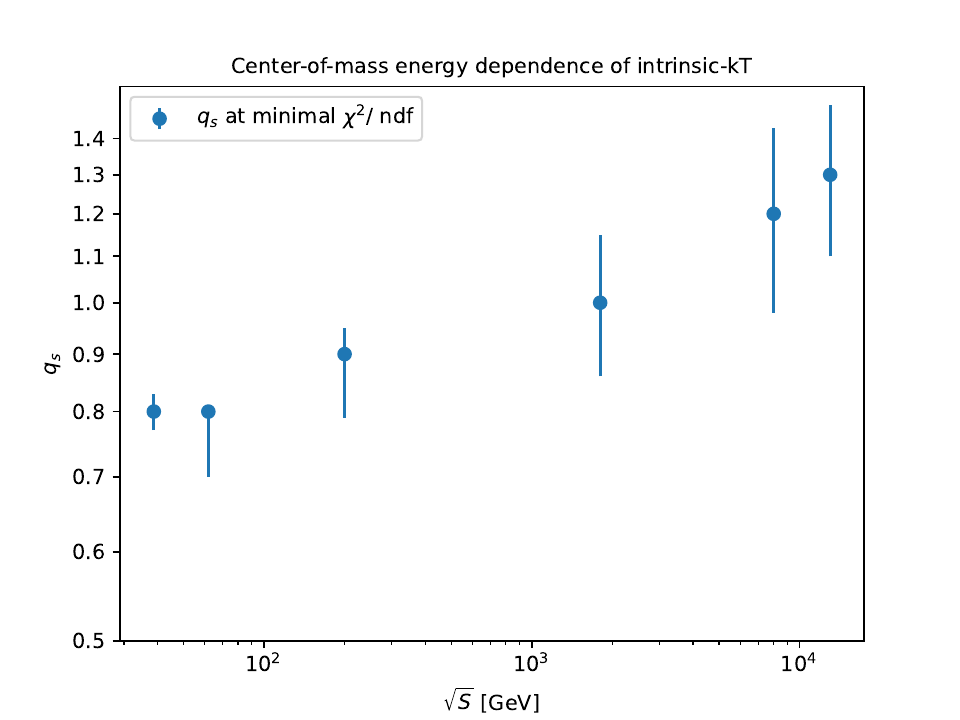}   
    \caption{
        Partonic branching phase space diagram (left, adapted
        from Ref.~\protect\cite{Hautmann:2019biw})
        and
        intrinsic $k_T$ parameter extracted from fits of PB TMD to DY $p_T$
        using dynamic soft-gluon resolution
        scale~\protect\cite{Hautmann:2025fkw} with $q_0 = 1$\,GeV (right).
    }
    \label{fig2}
\end{figure}

\paragraph{Soft gluons at low $k_T$}

We analyze the difference between the two panels of Fig.~\ref{fig:extract-intrins}
in terms of the partonic branching phase space~\cite{Hautmann:2019biw} shown in
Fig.~\ref{fig2} (left).
In this diagram, the vertical axis represents the transverse momentum $q_T$ of the
emitted parton, with hard emissions at the top.
Along this axis, emissions with $q_\perp<q_0\sim 1$\,GeV are considered soft and non
resolvable.
The horizontal axis corresponds to the longitudinal momentum fraction $z$ carried by
the emitting parton after a splitting, which is kinematically limited by the energy
scale at the branching $\mu$ (diagonal lines).
The evolution is carried out between the hard scale $\mu$ and the small scale
$\mu_0\sim1$\,GeV at which non-perturbative structure of the hadron is parameterized.

An essential feature of the PB TMD is its treatment of soft, non resolvable branchings
(gray region in the diagram), which corresponds to non-perturbative Sudakov effects.
This region dominates the PB TMD results for the rapidity evolution 
kernel~\cite{Martinez:2024mou,Martinez:2024twn} 
at large transverse distances $b$. 
Such results may be compared with analogous ones from fits of DY
data~\cite{Moos:2025sal,Bacchetta:2024qre,Bury:2022czx} 
and from lattice 
calculations~\cite{Bollweg:2025iol,Bollweg:2025ecn,Avkhadiev:2024mgd,LatticePartonLPC:2022eev,Deng:2022gzi,LatticePartonLPC:2023pdv} 
(in particular, regarding the asymptotic large-$b$ 
behavior~\cite{Collins:2014jpa,Hautmann:2020cyp,Boglione:2023duo,Hautmann:2007cx}).

The dependence of PB results in the treatment of the soft, non 
resolvable region is highlighted in Ref.~\cite{Hautmann:2025fkw} by
stopping the evolution at the ``dynamical'' soft-gluon resolution scale 
$z_\text{dyn}$~\cite{Hautmann:2019biw} that corresponds to the red line
in Fig.~\ref{fig2} (left).
First, it is verified by an \texttt{xFitter}~\cite{HERAFitter}
analysis that  precision deep inelastic scattering (DIS) measurements
are well described by applying the soft-gluon resolution 
$z_\text{dyn}$ in the TMD evolution.
Second, the analysis of Fig.~\ref{fig:extract-intrins} is repeated to
extract the parameter controlling the intrinsic $k_T$ behavior.
The results, shown in Fig.~\ref{fig2} (right), exhibit an increase in
the value of the extracted intrinsic $k_T$ with energy, similar to
\textsc{Herwig} and \textsc{Pythia}.
A similar study~\cite{Bubanja:2024puv} demonstrates that the slope
increases with larger $q_0$.

\paragraph{Correlated uncertainties and the  forward region}

As mentioned in the introduction, one can see in Fig.~\ref{fig:extract-intrins} (right)
that the PB TMD fits to the LHCb measurements~\cite{LHCb:2021huf}, as well as to
some of the lower-energy measurements, favor lower values of intrinsic $k_T$.
This raises the question of possible physical effects which may be responsible 
for this: e.g., the impact of 
large-$x$ parton distributions in the forward LHCb region 
and at low energy, the TMD flavor dependence in 
the valence and sea quark sectors, or
the interplay of large-$x$ and small-$x$ effects on transverse 
momentum in the  forward-rapidity asymmetric kinematics.

\begin{figure}[h]
    \centering
    \includegraphics[width=0.5\textwidth]{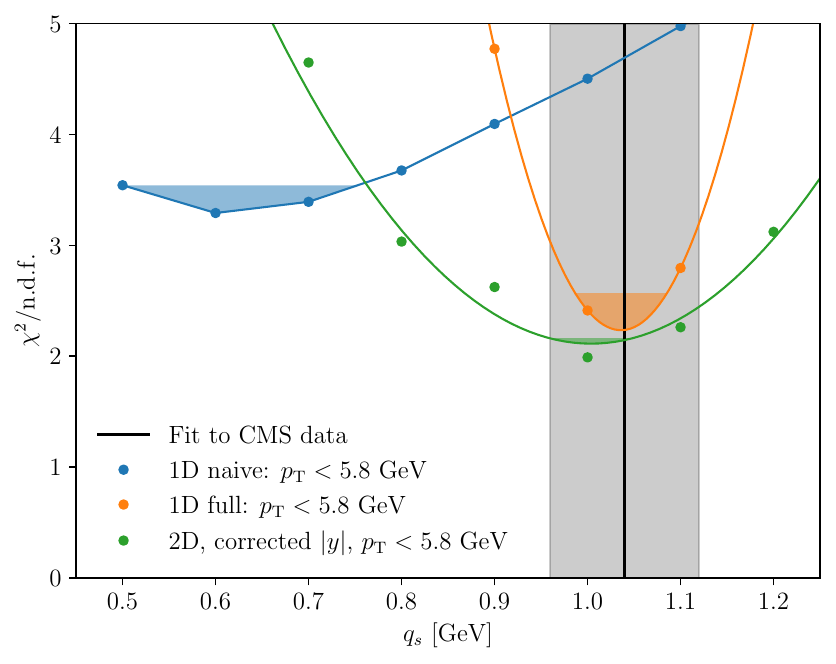}\relax
    \includegraphics[width=0.4\textwidth]{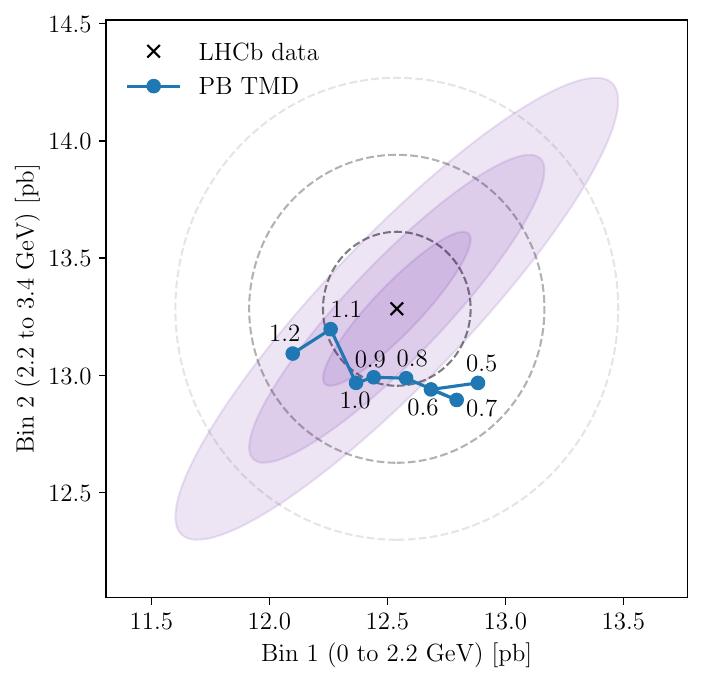}
    \caption{%
        (Left)  intrinsic-$k_T$  values  extracted from fits 
        to LHCb data: (blue) no correlations; (orange) including correlations;
        (green) including correlations and correcting the rapidity distribution.
        (Right) the effect of including correlations in the case of the 
        LHCb dataset~\protect\cite{LHCb:2021huf}: 
        dashed lines corresponds to no correlations, shaded areas correspond to 
        correct correlations included.
    }
    \label{fig-lhcb}
\end{figure}

In Ref.~\cite{Favart:2025inprep} we focus on the LHCb measurement.
The approach followed for the result shown in Fig.~\ref{fig:extract-intrins} (right)
is based on minimizing the $\chi^2$ between the measurement and the predicted
$p_T$ distributions, neglecting correlations among uncertainties
in the measurement made in the different bins.
We start by qualitatively reproducing these results (blue line in Fig.~\ref{fig-lhcb})
using the first four bins of the $p_T$ distribution.
We then make two changes to obtain the orange line: we scale the prediction to match
the measured total cross section, and we take correlations into account.
Finally, we fit the 2D measured $p_T$-rapidity distribution to obtain the green curve;
for this result, we include a correction of the inclusive rapidity distribution 
predicted by the PB TMD.
We make two observations:
a) in our fits, the minimum is shifted to a value compatible with the main result
   of Ref.~\cite{Bubanja:2023nrd}
   extracted from the CMS DY measurement~\cite{CMS:2022ubq}, and
b) the minimum is steeper, resulting in a more precise determination of the
   intrinsic $k_T$ parameter.

The reason for these changes is illustrated in the right panel of
Fig.~\ref{fig-lhcb}.
The LHCb measurement is projected to the plane formed by the first two
bins of the $p_T$ distribution.
The PB TMD prediction for different values of the intrinsic $k_T$ parameter is
represented as a blue line.
The experimental uncertainty, accounting for correlations, is shown as the 
diagonal ellipses for $\chi^2=1$, 2, and 3.
The dashed lines show the correspond to the same uncertainty without accounting
for the correlation between the bins.
Without the correlation, the $\chi^2<1$ region includes values of $q_s$ between
0.8 and 1.1\,GeV.
Accounting for the correlation reduces this to a fraction of the interval
between 1.0 and 1.1\,GeV.
Taking correlations into account is thus essential to fully exploit the
constraining power of a measurement.

\paragraph{Summary and outlook}

We presented two effects influencing extractions of the intrinsic $k_T$
parameters of Monte Carlo generators.
We first illustrated that the treatment of soft, non resolvable gluon
emissions has an impact on the center-of-mass-energy dependence of the
extracted parameters.
Then, we turned to experimental effects and highlighted the importance of a
correct treatment of correlated uncertainties in fits.
In Ref.~\cite{Favart:2025inprep}, we find that other effects are important
for the description of the LHCb data,
including
the treatment of QED final-state radiation and background processes.
Once these effects are taken into account, good sensitivity to
nonperturbative TMD dynamics can be achieved in the forward region.

\paragraph{Acknowledgments}
We thank the organizers 
of EPS-HEP2025 for the excellent conference and 
the kind invitation. We acknowledge useful 
conversations with A.~Bermudez Martinez, 
H.~Jung, H.~Li, M.~Mangano, N.~Raicevic 
and H.~Yin. 
L. Moureaux acknowledges support by the DFG under Germany’s Excellence Strategy 390833306 -- EXC 2121: Quantum Universe.
A. Lelek acknowledges funding by Research Foundation-Flanders (FWO) (1278325N).
L. Favart is supported by the ``Fonds de la Recherche Scientifique -- FNRS''.

\setlength{\bibsep}{5pt plus 0.3ex}

\end{document}